\documentclass{article}
\usepackage[utf8]{inputenc}

\usepackage[final]{nips_2017}

\usepackage[utf8]{inputenc} 
\usepackage[T1]{fontenc}    
\usepackage{hyperref}       
\usepackage{url}            
\usepackage{booktabs}       
\usepackage{amsfonts}       
\usepackage{nicefrac}       
\usepackage{microtype}      

\title{Taking Ethics, Fairness, and Bias Seriously in Machine Learning for Disaster Risk Management}

\author{
  Robert ~Soden\\
  Columbia University\\
  New York, NY 15213 \\
  \texttt{robert.soden@columbia.edu} \\
  \And
  Dennis Wageanar \\
  Deltares \\
  Netherlands \\
  \texttt{dennis.wagenaar@deltares.nl} \\
   \AND
  Dave Luo \\
  World Bank \\
  United States \\
  \texttt{dave@anthropo.co} \\
  \AND
  Annegien Tijssen \\
  Deltares \\
  Netherlands \\
  \texttt{annegien.tijssen@deltares.nl} \\
}

\begin{document}

\maketitle

\section{Introduction}

This paper highlights an important, if under-examined, set of questions about the deployment of machine learning technologies in the field of disaster risk management (DRM). While emerging tools show promising capacity to support scientific efforts to better understand and mitigate the threats posed by disasters and climate change, our field must undertake a much more careful assessment of the potential negative impacts that machine learning technologies may create. We also argue that attention to these issues in the context of machine learning affords the opportunity to have discussions about potential ethics, bias, and fairness concerns within disaster data more broadly. In what follows, we first describe some of the uses and potential benefits of machine-learning technology in disaster risk management. We then draw on research from other fields to speculate about potential negative impacts. Finally, we outline a research agenda for how our disaster risk management can begin to take these issues seriously and ensure that deployments of machine-learning tools are conducted in a responsible and beneficial manner.
 
Disasters triggered by natural hazards such as earthquakes or tropical storms are a major development challenge, and their risks are increasing as a result of climate change, human settlement patterns, and other social and political factors. Disaster are modelled for risk insurance products [9], infrastructure planning [19] and emergency management [2]. In the case of disaster insurance, a premium is paid in return for a compensation in case of a disaster and risk models are used to set this premium. In infrastructure planning, risk models are applied to assess the feasibility of specific risk reduction infrastructural investments [19], or help derive a robust long-term disaster risk management strategy [1]. They can also be applied to optimize the dimensions for infrastructural measures [11], help with spatial planning [4,5] or screen for places for further investigation that now or in the future may have a large disaster risk [20]. Disasters are also modelled for emergency management for example in forecasting or near real time warning systems [2]. This can be applied to issue warnings, make better informed decisions or prioritize humanitarian aid.

Machine learning holds the potential to help with these applications, especially when coupled with computer vision and geospatial technologies, by providing more accurate or lower-cost impact estimations based on improving the underlying hazard or vulnerability models. These methods may in the future automate and improve the interpretation of remote sensing hazard data, make detailed exposure data including many building characteristics available from interpreting aerial and/or street view data and combine it all with models trained on historical damage records. This may lead to much more accurate models than we have today that may be applied in new ways. It may in some cases also introduce cheaper lower quality models that are fully data-driven, or lacking in human oversight. Experiments and early trials of these technologies are already being undertaken by the World Bank’s Global Facility for Disaster Reduction and Recovery (GFDRR) and other development agencies in a number of locations around the world [6] to accomplish tasks such as:

\begin{itemize}
\item Rapidly evaluate large amounts of satellite and radar imagery to understand the extent of an area affected by flooding 
\item process street-level photography of building stock and other infrastructure to predict damage caused by hurricanes or earthquakes of a given intensity, OR
\item assess long-term urban growth patterns to gain in depth understanding of potential future vulnerabilities to disaster
\end{itemize}

\section{Concerns}

While taking into account the potential benefits of machine-learning tools to disaster risk management, we urgently need to develop a better understanding of the potential for negative, unintended consequences of their use. Significant attention is currently being given by academics, journalists, and the public to questions of the ethics and bias of machine-learning systems across a variety of domains including facial recognition [10], automated weaponry [18], search engines [13], and criminal justice [7]. Despite similar potential for negative impacts of these tools in disaster risk management, our community has not given these issues as much attention as other fields. Specific threats that machine-learning technologies present in this space include:

\begin{itemize}
\item Perpetuating and aggravating societal inequalities through use of biased training datasets
\item Aggravating privacy and security concerns in Fragility, Conflict and Violence (FCV) settings through combination of previously distinct datasets
\item Limiting opportunities for public participation in disaster risk management due to increased complexity of data products
\item Reducing the role of expert judgement is data and modeling tasks in turn increasing probability of error or misuse
\item In addition, many systems do not adequately communicate their methods or degrees of uncertainty, which increases the chance of misuse.
\end{itemize}
 
Each of these issues has already been documented in other domains and is worth examining the field of disaster risk management.  These are concerns that need to be weighed seriously against the potential benefits before introducing new technologies into disaster risk management information systems. A number of technology companies\footnote{e.g. https://www.microsoft.com/en-us/ai/our-approach-to-ai} and research institutions\footnote{e.g https://ainowinstitute.org/aap-toolkit.pdf} have developed guidelines for evaluating machine-learning systems but this work is still evolving. In some cases, like facial recognition, experts have begun to recommend not using it all and they have been banned in a number of jurisdictions in the United States\footnote{For example, the City Councils of both Oakland and San Francisco voted to ban the use of facial recognition technologies in 2019. These cases are especially noteworthy given the high concentration of technology experts living and working in these cities.}. It is too early to know how this debate will play out in the field of disaster risk management so it is worth proceeding with caution.
 
In addition, the attention given to risks of ML create the opportunity to explore how existing and widely-used disaster data tools like risk modeling or damage assessment pose very similar concerns that have for too long gone unexamined. All disaster data is limited, and provides a necessarily incomplete view of the complex phenomena it is meant to describe [3,12,17]. Too often we measure what we have data for, or what is possible to measure, rather than what matters most. The increased attention to questions of ethics and bias in ML systems more broadly might serve as an opportunity to drive conversations in our field about the limits of disaster data more generally. Many of the sources of bias or ethical concerns in machine-learning systems originate in, or share common roots with other kinds of data used to understand disaster risks and impacts. This includes issues such as 1) property values determining priority areas for protection, 2) the neglect of areas with poor or missing data (often also linked to lack of resources), 3) privacy concerns (which may be aggravated by ML and other big data techniques), 4) how the lack of gender and age disaggregated data on disaster risk masks differential vulnerabilities, and 5) the importance of public participation and the voice of residents of areas portrayed by models as "at risk".
 
\section{Recommendations}
In order for machine learning technologies to be deployed in the disaster risk management context in a responsible manner, the community of experts and practitioners working on these tools urgently need to take questions of ethics, bias, and fairness seriously. We recommend that the following actions be taken:
\begin{enumerate}
\item Proceed with caution and conduct threat assessments of all new applications of AI or ML technologies. Recognize that, as a community, we haven’t yet conducted due diligence around the potential unintended harms that these tools may cause.
\item Take guidance from existing guidelines from existing recommendations related to data in the development and humanitarian sector including the Principles of the Open Data for Resilience Initiative [15], the Signal Code by the Harvard Humanitarian Initiative [8], or the Principles for Digital Development Initiative\footnote{https://digitalprinciples.org}.
\item Convene discussions and meetings cross-organization to share knowledge, develop guidelines for evaluation and deployment of machine learning tools in the disaster risk management context. The authors of this paper, representing Columbia University, Deltares, and the World Bank have already begun discussions towards this end and are actively recruiting participants from other organizations.
\item Learn from the experiences of other fields and domains. While the conversation about ethical use of machine learning in disaster risk management is nascent, there are numerous studies and cautionary examples from other contexts that we can draw on when evaluating the potential consequences of these technologies.
\item Detailed analysis of specific cases [eg 16] is needed urgently to make further progress in understanding the ethical and political consequences of the design choices embedded into the information systems we use to understand disaster.
\item Work in transparent fashion, in collaboration with communities and people who are represented in/by these technologies. Where possible and appropriate support open-source and open data approaches.
\end{enumerate}

\section*{References}
[1] Aerts, J.C.J.H., Botzen, W.J.W., Emanuel, K., Lin, N., de Moel, H. (2014). Evaluating Flood Resilience Strategies for Coastal Megacities. Science 344 (6183), 473-475.

[2] Coughlan de Perez, E., van den Hurk, B., van Aalst, M. K., Jongman, B., Klose, T., and Suarez, P.: Forecast-based financing: an approach for catalyzing humanitarian action based on extreme weather and climate forecasts, Nat. Hazards Earth Syst. Sci., 15, 895-904, https://doi.org/10.5194/nhess-15-895-2015, 2015.

[3] Crawford, K. and Finn, M., 2015. The limits of crisis data: analytical and ethical challenges of using social and mobile data to understand disasters. GeoJournal, 80(4), pp.491-502.

[4] De Bruijn, K. M., Klijn, F., van de Pas, B., and Slager, C. T. J.: Flood fatality hazard and flood damage hazard: combining multiple hazard characteristics into meaningful maps for spatial planning, Nat. Hazards Earth Syst. Sci., 15, 1297-1309, https://doi.org/10.5194/nhess-15-1297-2015, 2015.

[5] De Bruijn, K.M., Klijn, F., 2009. Risky places in the Netherlands: a first approximation for floods. Journal of Flood Risk Management volume 2, issue 1, March 2009, Pages 58-67.

[6] Deparday, V., Gevaert, C.M., Molinario, G., Soden, R. and Balog-Way, S., 2019. Machine Learning for Disaster Risk Management. World Bank Publications.
 
[7] Eubanks, V., 2018. Automating inequality: How high-tech tools profile, police, and punish the poor. St. Martin's Press.

[8] Greenwood, F., Howarth, C., Poole, D., Raymond, N. and Scarnecchia, D., 2017. The Signal Code: A Human Rights Approach to Information During Crisis. Harvard, MA.

[9] Grossi, P.,  Kunreuther, H., 2005. Catastrophe Modeling: A new approach to managing risk. Heubner International Series on Risk, Insurance and Economic security. ISBN: 0-387-23082-3.

[10] Keyes, O., 2018. The misgendering machines: Trans/HCI implications of automatic gender recognition. Proceedings of the ACM on Human-Computer Interaction, 2(CSCW), p.88.

[11] Kind, J.M.(2013), Economically efficient flood protection standards for the Netherlands. Journal of Flood Risk Management, 7(2), 103-117, 2013.

[12] Liboiron, M., 2015. Disaster Data, Data Activism: Grassroots Responses to Representing Superstorm Sandy. In Extreme weather and global media (pp. 144-162). Routledge.
 
[13] Noble, S.U., 2018. Algorithms of oppression: How search engines reinforce racism. nyu Press.
 
[14] Reisman, D., Schultz, J., Crawford, K., Whittaker, M. 2018. Algorithmic Impact Assessments: A Practical Framework For Public Agency Accountability. AI Now.

[15] Simpson, A., Deparday, V., Dinethi, S. Soden, R. 2016. Open data for resilience initiative: policy note and principles. World Bank Publications.
 
[16] Soden, R. and Kauffman, N., 2019, April. Infrastructuring the Imaginary: How Sea-Level Rise Comes to Matter in the San Francisco Bay Area. In Proceedings of the 2019 CHI Conference on Human Factors in Computing Systems (p. 286). ACM.
 
[17] Soden, R. and Palen, L., 2018. Informating crisis: Expanding critical perspectives in crisis informatics. Proceedings of the ACM on Human-Computer Interaction, 2(CSCW), p.162.
 
[18] Suchman, L.A. and Weber, J., 2016. Human-machine autonomies.
Van der Most, H., Tanczos, I., De Bruijn, K.M. and Wagenaar, D.J.,(2014), New, Risk-Based standards for flood protection in the Netherlands. 6th International Conference on Flood Management (ICFM6), September 2014, Sao Paulo, Brazil. 2014.

[19] Wagenaar, D.J., Dahm, R.J., Diermanse, F.L.M., Dias, W.P.S., Dissanayake, D.M.S.S., Vajja, H.P, Gehrels, J.C., Bouwer, L.M., 2019. Evaluating adaptation measures for reducing flood risk: A case study in the city of Colombo, Sri Lanka. International Journal of Disaster Risk Reduction. Volume 37, July 2019, 101162.

[20] World Bank, 2018. Afghanistan—Multi-hazard risk assessment. Washington, DC. https://www.gfdrr.org/en/publication/afghanistan-multi-hazard-risk-assessment?deliveryName=DM7618.

\end{document}